# Artificial intelligence and machine learning applications for cultured meat

**Michael E. Todhunter**[1†#], **Sheikh Jubair**[2†#], **Ruchika Verma**[2], **Rikard Saqe**[3], **Kevin Shen**[3], **Breanna Duffy**[4*†]

[1]Todhunter Scientifics, Minneapolis, MN, USA

[2]Alberta Machine Intelligence Institute, Edmonton, AB, Canada

[3]University of Waterloo, Waterloo, ON, Canada

[4]New Harvest, Sacramento, CA, USA

**\* Correspondence:**
Breanna Duffy
breanna@new-harvest.org

[†] These authors contributed equally to this work

[#]These authors share first authorship

**Abstract**

Cultured meat has the potential to provide a complementary meat industry with reduced environmental, ethical, and health impacts. However, major technological challenges remain which require time- and resource-intensive research and development efforts. Machine learning has the potential to accelerate cultured meat technology by streamlining experiments, predicting optimal results, and reducing experimentation time and resources. However, the use of machine learning in cultured meat is in its infancy. This review covers the work available to date on the use of machine learning in cultured meat and explores future possibilities. We address four major areas of cultured meat research and development: establishing cell lines, cell culture media design, microscopy and image analysis, and bioprocessing and food processing optimization. This review aims to provide the foundation necessary for both cultured meat and machine learning scientists to identify research opportunities at the intersection between cultured meat and machine learning.

## 1 Introduction

Food production generates about a quarter of global greenhouse gas emissions and causes other negative impacts on the environment and human health (1,2). Animal products, including meat, seafood, eggs, and dairy contribute more than 56% of food's emissions, despite providing only 37% of protein and 18% of calorie intake (3). Meat production is a large sector, producing 328 Mt in 2020 and expected to expand to 374 Mt by 2023, based on estimates from the Organisation for Economic Co-operation and Development (OECD) and the Food and Agriculture Organization (FAO) of the United Nations (4). In light of projected growth in the global population and income, these estimates projected that meat consumption will increase by 14%. To meet global meat demand and limit global warming to 1.5°C there is a need for major changes in the production of meat (5,6). Cultured meat

(CM), known by many names including "cell-based" or "cultivated" meat, is an emerging technology that uses tissue engineering and biomanufacturing techniques to produce animal meat through cell culture rather than animal husbandry. Proponents of the technology herald its potential to provide an option for producing animal agriculture products with reduced environmental, ethical, and health impacts (7,8). However, major technological challenges remain in bringing CM products to market and achieving their proposed benefits (9). Many challenges stem from the fact that the technologies for mammalian tissue culture come primarily from the medical field, where the scale is much lower and the market has weaker incentives to reduce the costs of production. However, if these technologies are to be applied to food, the challenges of scale and cost must be addressed.

Some specific improvements to CM can be, and have been, made using traditional experimental approaches. However, tackling some of the more complex research questions requires more advanced approaches and experimental conditions. In recent years, an increasing number of groups have been using methods enhanced by artificial intelligence (AI), and in particular its subset machine learning (ML), for such tasks. ML can streamline experiments, predict optimal results, and reduce experimentation time and resources. There are many opportunities for ML to accelerate research development and reduce costs in CM. Some companies have indicated the use of ML in CM product development or CM-associated services (10–17), but very little of this progress in applying ML to CM has been shown or validated in the public domain. Academic and government-supported research in this space is emerging, including at the University of California at Davis, Virginia Tech, Tufts University, and The CentRe of Innovation for Sustainable banking and Production of cultivated Meats (CRISP Meats). However, few publications on the topic exist to date (18–21). An increase in open public research on the use of ML to optimize and scale CM production would greatly accelerate the application of ML to the CM field.

In this review, we aim to provide a foundation necessary for researchers, from the CM or ML fields, to identify research opportunities at the intersection between CM and ML. We first provide a brief overview of both fields. Subsequent sections delve into both existing and potential ML applications for optimizing cell lines, formulating culture media, aiding cell culture microscopy and image analysis, and optimizing bioreactor and food processing parameters. Note that we have focused on CM challenges to which ML can be applied, and this should not be seen as a comprehensive review of all challenges in the CM or ML fields. As applications of ML in CM are limited, we discuss how ML methods that have been applied in other areas of bioinformatics can be adopted to solve tasks in CM.

## 2    Background on the Fields of Cultured Meat and Machine Learning

### 2.1    Cultured Meat

CM aims to replicate the taste and texture of animal tissue within a manufacturing system using animal cells (Figure 1). First, cells from a species of interest are selected or engineered for desirable growth and differentiation traits in vitro. The selected cells are grown in a suitable medium, providing the nutrients and signaling cues that would normally be provided in the body. At early stages, such as cell selection, the cell culture may start at a small scale in plastic dishes or flasks. Cell culture is eventually scaled up to industrial-scale bioreactors, devices capable of controlling environmental temperature, pH, dissolved oxygen, and nutrient exchange at large volumes (22). Once enough cells are grown, they are differentiated into mature cell types. At this stage, the cell culture may be formed into a tissue (i.e. structured product, such as a steak) or a cell slurry, which is later processed into a meat product (i.e. unstructured product, such as ground beef).

This process leans heavily on medical tissue engineering, an area of research that has been studied for nearly 40 years and has been commercialized at a small scale for simple grafts of cell-laden scaffolds for skin and cartilage. However, complete tissues, such as skin with hair follicles and sebaceous glands or functional muscle, still face technological barriers (23). While tissue function is not critical for CM, using this technology for the production of food comes with unique constraints, especially the need for larger scale, lower costs, and materials that are both edible and palatable.

Commercial and academic interest in CM has grown rapidly in the last decade. The number of companies working on CM grew from 1 to over 170 from 2011 to 2023, with over 3 billion dollars of investment (24). Similarly, academic interest in the field has grown dramatically, with 350+ papers published on CM in the last 2 years, more than all other years prior (with tracking) combined (24). In parallel, cost estimates for CM have come down significantly in the last decade: from the first CM demonstration in 2013 of a 140-gram burger at approximately €250,000 (25), to recent claims of costs as low as $7.70/lb from industry developers (26). However, these industry cost claims have yet to be proven publicly. Furthermore, CM production has yet to be shown on a scale close to that needed to offset even a fraction of current meat consumption (27). Current production capacity is not known, but estimates range from 1-10 kg/y, compared to the $3.2 \times 10^{11}$ kg/y produced by conventional meat (28). Highly optimized industrial mammalian cell lines, such as Chinese hamster ovary (CHO), are still produced at a much lower scale and higher cost than needed for food production (28). Given that consumers are unlikely to want to consume hamster ovary cells, significant technological challenges must be overcome for meat and seafood-relevant cells to be produced at a scale and cost needed to replace a meaningful portion of the conventional meat market.

## 2.2 Machine Learning

An ML workflow involves a series of steps, as illustrated in Figure 2, which starts with preparation of the dataset. A dataset typically consists of datapoints, each one an observation with features that describe the datapoint. The type of data varies widely: numerical, time series, text, images, audio, video, sequential, graph, or any combination of these. Since the data must be in a numerical form for the subsequent steps, a data transformation step converts the data to numerical values. The next step is the preprocessing of raw data, where procedures such as imputing missing values and reducing data dimensionality are undertaken.

Subsequently, an integral component of the ML methodology is dividing data into distinct subsets: training, validation, and testing. Some common approaches, such as K-fold cross-validation, leave-one-out, and holdout validation, can typically be used for most problems (30). Prior to model training, it may be necessary to employ feature selection or extraction techniques on the training set to identify the most informative variables in the data, and these features must be used in both the test and validation sets to maintain the integrity of the evaluation process.

The culmination of this preparatory work is a dataset suitable for training an ML model. Typically, experiments are conducted with different ML algorithms/architectures (such as random forests, k-means, and deep neural networks) to settle on the most performant overall model. However, sometimes the choice of ML model depends on the type of data. For example, convolutional neural networks (CNNs) are preferred for image data since they can extract spatial features from the image (31). On the other hand, recurrent neural networks (RNNs) are typically used for sequential data since they can remember information in a long sequence through their gate mechanism (32).

The training set is instrumental in building the model since the model adjusts its parameters to learn the distribution of the training set. The validation set is required for fine-tuning the model's hyperparameters and ensuring that the model generalizes well beyond the training data, thereby avoiding overfitting. Finally, the test set provides a measure of the model's predictive accuracy and overall performance in real-world scenarios. Some of the typical performance metrics for classification tasks are accuracy, precision, recall, and F1-Score. Typical performance metrics for regression are $R^2$ score, mean absolute error, and mean square error (33). However, lots of other performance metrics are used depending on the tasks and types of data. The purpose of performance metrics is to compare the performance between different models and also to understand how a model is performing overall for the specific task.

**Types of Machine Learning:**

ML methodologies can be broadly classified into three categories: supervised, unsupervised, and reinforcement learning (RL). In supervised learning, the model is trained on a labeled dataset, where each example is paired with an outcome or label that aligns with the objective of the specific task at hand. Consider the goal of predicting gene expression levels from DNA sequences; here, the data point would be the DNA sequence, while the associated expression level serves as the label. The model hones its predictive capability by learning the relationship between input features and their corresponding labels. When it comes to evaluation, the trained model is tasked with predicting labels for new, unseen data points. Key traditional supervised learning models include logistic/linear Regression, k-nearest neighbours, and support vector machines. The supervised ML approach has made significant contributions across various domains in bioinformatics, enabling advancements in DNA segmentation, gene expression prediction, and protein structure prediction (34).

Unsupervised learning, in contrast, does not rely on labeled data. It is particularly useful when the goal is to unearth underlying patterns or structures within the data, independent of predefined outcomes. This makes unsupervised learning a potent tool for exploratory analysis, especially in scenarios where labeled data is scarce or when the structure of the data is not fully understood. Some of the key unsupervised learning approaches are k-means, hierarchical clustering, and density-based spatial clustering of applications with noise. Unsupervised learning has been successfully applied in grouping functionality-related genes, microarray analysis, and biological image segmentation (34,35).

RL is a dynamic and adaptive approach well-suited for situations where a machine is required to make a series of decisions to achieve a desired goal or perform an optimization task, offering a framework for learning through interaction (36). Within this paradigm, an agent—often an advanced ML model in deep RL —engages in a sequential decision-making process, each time interacting with a complex environment represented by a set of variables that define the current state of the environment. The agent executes actions, transitioning between states, and ultimately may reach a terminal state, signaling the conclusion of the decision sequence. The RL framework incorporates a system of rewards and penalties, with the agent receiving feedback in the form of rewards for beneficial actions or penalties for undesirable outcomes. The objective for the agent is to devise a strategy that maximizes cumulative rewards, thus steering towards the most optimal actions to attain its final goal. RL has recently been applied to bioinformatics (37–39) and gained lots of attention because of its success in areas, such as sequence alignment (40) and protein loop sampling (41).

**Neural Networks:**

Recent advances in applying ML to biology are based on neural networks, a subset of ML methods that can be employed in all three ML categories: supervised, unsupervised, and RL (42). ML models with neural networks are often referred to as deep neural networks when there are multiple layers of neural networks in the architecture of the model, a method more broadly known as deep learning. These neural network layers typically attempt to mimic the activity of brain neurons, where each neuron employs a mathematical function that alters the data it receives from the previous layer (43,44). At first, the data is fed into an input layer, which then connects to hidden layer(s) (used for computing), and finally an output layer, designed to deliver the final prediction. The model learns by optimizing the function parameters of each node. Some popular neural network architectures are feed forward networks, CNNs, RNNs, and transformers (43–45). Deep learning models typically capture complex biological processes and incorporate heterogeneous data in the model through its different layers, which may be a necessity for many optimization and prediction tasks in CM (46). It can be applied to both supervised and unsupervised scenarios. In RL, deep learning models can be employed as agents as well (47).

**Generative AI:**

Generative AI is another subfield of unsupervised learning which typically aims to generate new data or samples based on the patterns learned from the training data. Variational Autoencoders (VAEs) (48–50), which employ deep learning, are the first generation of generative AI models. VAEs typically employ two deep learning models: i) an encoder that encodes the input into a latent space and ii) a decoder that predicts the input by matching the distribution of the latent space by applying variational inference techniques. Together, the encoder and decoder minimize the reconstruction loss of the input. Figure 3 shows the general architecture of VAE.

The second generation of generative models is deep adversarial networks (51,52). These architectures also have two networks: a generator and a discriminator. Initially, generators generate new data randomly and discriminators classify whether the generated sample is original or generated. Iteratively, both the generator and discriminator learn how to generate a better sample and distinguish between original or generated sample. Eventually, the generator learns to generate realistic data samples that are difficult for the discriminator to distinguish from real data.

Recent advances in generative AI are mostly based on transformers (45), which have revolutionized various fields including natural language processing, computer vision, and bioinformatics (53,54). Transformers have demonstrated exceptional capabilities in capturing long-range dependencies and modeling complex sequential data. In bioinformatics, methodologies heavily draw inspiration from NLP techniques due to the inherent similarities between biological sequences and natural language texts. By employing transformers, researchers are able to effectively model biological sequences such as DNA, RNA, and protein sequences, leading to significant advancements in tasks such as sequence generation, structure prediction, and drug discovery (38,53–56).

**Graph Neural Networks:**

Graph neural networks (GNNs) are another area of ML that works with graph structured data and biological networks, such as via protein interaction networks, gene coexpression networks, and metabolic networks (57,58). A biological network or biological graph typically is arranged in terms of nodes and edges, where the nodes are biological entities (i.e. genes, proteins, metabolites, etc) and the edges indicate how these entities relate to one another. GNNs are based on the principle of message passing, where each node in the graph aggregates all embeddings (messages) of its neighbour nodes and updates the weights of pooled messages through a neural network. There are three main tasks in graph structured data. The first one is link prediction between two biological entities. Examples of link prediction are predicting the interactions between proteins or predicting interactions between genes/regulatory elements (59,60). The second task is node functionality prediction. For example, predicting an unknown function of a protein based on the physical interactions between proteins in the protein interaction network (61). The third application of ML in network analysis is to classify sub-network functionality. Figure 4 shows different ML tasks in a graph or network structured data. Muzio et al. classified functions of subnetwork based on a molecule's toxicity or solubility (61). In addition, ML is also applied to obtain subnetwork embedding where the subnetwork is represented as a vector preserving the important information within the subnetwork in a numeric form. This embedding can facilitate further analysis of the subnetwork and is used for various downstream tasks (62).

## 3 Cell Lines

Meat consists of various cell types, predominantly muscle cells (approximately 90%), with fat and connective tissue cells accounting for the remaining 10% (63). Additionally, there are some vascular, neural, and tissue-resident immune cells present in small amounts (63–65). To provide the taste, texture, and aroma expected in meat, most CM development has focused on muscle, fat, and connective tissue production. Cells will also likely contribute to the nutritional properties of CM products, and cell optimization may be used to tailor the nutritional profile of CM (66,67).

The cell types used to produce CM range from lineage-committed progenitor cells (e.g. muscle satellite cells, myoblasts, or preadipocytes) to stem cells able to differentiate into a broader set of cells (e.g. mesenchymal stem cells, embryonic stem cells, or induced pluripotent stem cells) (64). Developers may use primary cells, meaning cells that are isolated directly from an animal. However, cell lines, which are established cultures of cells that have been selected and optimized, are ideal because they are more consistent, characterized, and reduce the use of animals in the supply chain. Furthermore, immortalized cell lines or pluripotent stem cells are of particular interest due to their ability to escape the typical limits on population doublings seen in most primary cells (i.e. Hayflick limit) (68,69). A detailed review of cells used in the production of CM can be found elsewhere (65).

Few CM-relevant cell lines are currently well characterized and commercially available, with most coming from model organisms used in biomedical research, such as mouse, rat, or zebrafish, and cell lines for many agricultural species still need to be developed (70). Cell lines for marine species, especially invertebrates such as mollusks or crustaceans, are especially underdeveloped, with only a few reported cell lines and not all are food-relevant species (71–80). For many species, there is a lack of basic knowledge of their physiology and the biochemistry required for in vitro culture or immortalization (70,81). The lack of knowledge of molecular and genetic markers as well as few species-specific antibodies available makes identifying and curating cell lines difficult for under-studied species. ML can help biologists analyze complex cellular data to assist with identifying ideal cell line populations and optimizing existing cell lines through gene perturbations.

### 3.1 Network analysis is a tool to model biological interactions

Optimization of cell lines is often challenging because it requires understanding the "state" of a cell or cell population (i.e. what the gene and protein networks are doing) and selecting or engineering for desired cell states. Measuring and predicting these states involves interpreting complex interactions between genes and proteins, identifying those that are important for specific qualities, and predicting how perturbations will affect the whole network. ML can be used to model this complex network through an area of study known as network analysis. Network analysis is used throughout sections 3 and 4, so a general overview is first presented here.

Most of the earlier works in network analysis, such as DeepWalk (82–84), LINE (85), Node2Vec (86,87) are inspired by NLP or community detection methods of social networks. These methods primarily use multiple random walks from a specific node to obtain a subnetwork. The objective is to obtain a vector embedding of the specific node by following a skip-gram strategy employed in NLP (88). The obtained embedding maximizes the probability of predicting the neighboring nodes obtained by random walks. This embedding can be used for most of the network analysis tasks described previously. However, these approaches often do not preserve the graph structure information which may lead to information loss.

In recent years, GNNs have been employed in graph-structured data and biological networks. While the objective remains the same for network analysis with GNN, the primary difference between the NLP approach and the GNN is that since GNNs are built for working with graph structured data, each node in the graph aggregates information from its neighbours while preserving the connectivity information in the GNN. GNNs have been implemented to predict protein interactions (89,90), molecular interactions (91,92), metabolite-disease associations (93) and obtain subnetwork embeddings (94) that can be used to identify the functions of a biological subnetwork. Generative deep learning strategies, such as VAEs and deep adversarial network-based models, are also trained on biological networks by employing GNNs (95,96). The most famous example to date is AlphaFold2, a deep learning GNN model that was employed on amino acid sequences to predict more than 200 million protein structures (38). This is a significant advance in the field of structural biology which can be used in protein design, drug target prediction, cell type identification, and antibody development. AlphaFold2 can play a big role in designing new proteins and understanding the physical relationships between proteins (97).

Network analysis can employ multiple types of biological data within a multi-omics setting. Since each omics technique typically captures a specific biological process, integrating multi-omics data can provide a holistic overview of the biological process. In a multi-omics ML approach, different ML models are employed on different types of data (such as genomics, transcriptomics, proteomics, and metabolomics) to obtain numerical representations. These numerical representations are then combined together to obtain a more informative representation used to predict final outputs. For example, MOGONET employs a GNN on mRNA expression, DNA methylation, and miRNA expression data to predict disease information (98). Similar datasets can also be used for CM to predict cell line features. More applications of multi-omics ML methods are discussed in section 3.2.

Furthermore, network analysis can help determine aroma and flavor (99). Since meat aroma and flavor are largely controlled by metabolic pathways (100), network analysis can be applied to biosynthetic pathways to enhance or add flavors to CM products. Network optimization has been used to design yeast that overexpress licorice glycoside (101), and although licorice is a flavor that few would want in a steak, one could imagine more savory corollaries. In other yeast experiments, transcriptomic network analysis has been used to improve acid resistance, and acid resistance is just as important to mammalian culture as to microbial culture (102).

## 3.2 Machine Learning Can Help to Analyze Complex Omics Data to Identify and Characterize New Cell Lines

"Omics" approaches, such as genomics, transcriptomics, proteomics, and metabolomics, are powerful tools for identifying and characterizing cell lines. In particular, RNA sequencing (RNA-seq) technologies are commonly used to quantify cellular gene expression to validate and optimize cell lines. However, analyzing RNA-seq or other -omics data across many candidate cells is a complex and daunting analytical task. ML can help these analyses in multiple ways, including by grouping functionality-related cells using an unsupervised approach (94), profiling gene expression using a supervised approach (103), and identifying different tissue types using unsupervised (104,105) and semi-supervised approaches (106).

Using gene expression data to cluster cells based on cell type or behavior can help explain heterogeneity among cell populations and discover subpopulations with beneficial characteristics. When establishing cells for CM production, scientists may want to isolate only certain cell types with optimal attributes or remove undesirable cell types. For example, using single cell RNA-seq (scRNA-seq) Messemer et al. found that an isolation from cattle muscle contained 11 distinct cell types (107). This work led to a better understanding of the cells derived from a primary isolation, as well as cell surface markers suitable for the identification and separation of populations by flow cytometry. Additionally, in a recent preprint, Melzener et al. used RNA-seq to study subpopulations during muscle differentiation, understanding cell fates with an aim to improve the efficiency of cell differentiation and maturation (108). ML is also increasingly being explored for the modeling of cell trajectories via scRNA-seq (109).

Unsupervised ML can help to map cellular heterogeneity by grouping functionality-related cells, identifying cell sub-populations, and performing dimensionality reduction (104,105,110). Typically, the input to the unsupervised ML model is gene expression data obtained from RNA-seq. In conventional ML frameworks, which are not based on neural networks, the outcome is generally an assigned cluster number for each cell or gene. For an in-depth exploration of how traditional ML techniques are applied in this context, we direct readers to the comprehensive review by Petegrosso et al. (111). In contrast, unsupervised deep learning methods predominantly leverage autoencoders, which compress high-dimensional cellular data into a more manageable lower-dimensional space while retaining essential information (104,105,110). This lower dimensional representation of the encoder can be used to obtain clusters of cells (104,105,112–114) and can be further fine tuned for gene expression profiling (42,115).

Recently, another family of autoencoders that uses GNNs has been employed to obtain the lower dimensional representation of transcriptomics data. These GNN-based autoencoders use the knowledge of biological networks, such as gene-gene relationships, cell interaction networks, protein interaction networks and biological pathways, along with gene expression data to obtain a more robust and informative representation of cells and genes (94,110,116–120). The GNN autoencoders are unique compared to traditional autoencoders as they encode information on biological interaction between entities along with structural information and biological properties.

Semi-supervised approaches have been employed in many tasks, such as learning responses due to gene perturbation, different functional score prediction due to gene knockouts, identifying different tissue types, and transcriptome analysis

(106,121–126). These models are adept at leveraging both labeled and unlabeled data, typically employing the unlabeled data to train an autoencoder and then using the labeled data to fine-tune the autoencoder toward the target outcomes (106,123,127). Fine-tuning is the process of adjusting pretrained model parameters for a specific task by utilizing a small labeled dataset. The semi-supervised approach is notably different from its unsupervised counterpart as it provides a degree of guided learning, which is crucial when the available labeled data is sparse but critical for the identification of specific labels. For instance, while the categorization of cells based on functional similarities may fall under unsupervised learning—grouping cells by inherent characteristics—the task of pinpointing cell doublets could benefit from a semi-supervised model that utilizes a limited set of known doublet samples to enhance its predictive accuracy.

In addition to transcriptomic data, further single-cell multimodal sequencing technologies have been developed that provide cell-specific information, such as chromatin accessibility (scATAC-seq) (128) and surface proteins (CITE-seq) (129). These data types provide complimentary insight into scRNA-seq, such as improving accuracy in modeling gene regulatory networks in the case of scATAC-seq (130,131). This understanding can aid in CM-related tasks, such as identifying transcriptomic/epigenetic markers predictive of high proliferation and differentiation potential given previously observed variability in primary cell culture performance (132–134).

In an ML context, autoencoders and GNN-based deep learning models are mostly applied to this multimodal data (135). Some of these autoencoders employ individual encoders and decoders for each modality. We refer the readers to a review of multimodal single-cell models, and a review on general best practices for single cell analysis, to learn more about the topic (135,136).

### 3.3 Antibody Design for Characterization and Isolation of Novel Species can be Aided by Machine Learning

In cell line development it is crucial to both characterize cells, commonly via omics data (as discussed above) or visual identification (discussed in section 5), and isolate the cells of interest, typically using flow cytometry cell sorting. Visual identification and flow cytometry both require the use of antibodies with specific binding affinity to known cellular markers. Markers of muscle cell differentiation are well understood for most mammalian species (137). However, in under-explored species, such as fish or aquatic invertebrates, these proteins are not always shared with mammalian cells or, to the degree they are, have low sequence conservation (138). For example, recent work with an Atlantic mackerel (Scomber scombrus) skeletal muscle cell line found that antibodies for the muscle satellite cell marker paired-box protein 7 (PAX7) was successful, while early myogenic marker myoblast determination protein 1 (MYOD) was not (80). Surface markers, which are most commonly used during flow cytometry cell sorting, are particularly understudied in CM-relevant cell types, with most existing research focused on immunology (138,139).

Attempts at validating commonly used mammalian antibodies in fish have led to issues with cross-reactivity and specificity, suggesting that establishing cell lines for CM research will require the repurposing of existing antibodies, if not the development of fully novel antibodies (140). The use of ML on omics data has been shown to aid antibody development, such as via improving species cross-reactivity, antibody co-optimization, and binding affinity (141–143). The autoencoder model totalVI, trained on joint scRNA-seq and CITE-seq data, is able to provide insight into many key variables that would aid antibody development: the identification of novel differentially expressed features to target in a given cell subpopulation, the improved prediction of false positive/negative surface proteins, the reduction of technical bias common in antibody-based measurements such as "background",  and the improvement of experimental design by helping determine optimal antibody titrations/sequencing depths for balancing cost and signal-to-noise ratio (144). For a full review of computational methods relevant to antibody development, readers are pointed to (145).

### 3.4 Machine Learning Can Map and Enhance Genetic Traits to Optimize Cell Lines

Gene editing can be used to enhance or alter cellular traits to generate cell lines optimized for CM production, such as accelerating growth, extending growth (i.e. immortalization), reducing input costs, or tailoring the flavor and nutrition. For example, manipulating cellular metabolism has a large potential for increasing the efficiency (and therefore reducing costs) of CM products (9,146). As another example, Stout et al. engineered cells to overexpress FGF2, eliminating the need for FGF2 supplementation in media through autocrine signaling (147). Because many of the species used in CM are under-studied, their genome regulatory networks are not yet well understood, slowing efforts for gene editing. ML offers an opportunity to accelerate gene editing technology.

Genes typically have many regulatory regions, such as upstream and downstream regions, untranslated regions, promoters, and enhancers. These regulatory regions determine where, when, and how much a gene is expressed. Thus, to apply gene editing technology, identifying these regulatory regions is an essential task where ML models have been applied successfully (53,148,149). The most relevant modern ML techniques demonstrated to be useful for these tasks are generative adversarial networks (150,151) and convolutional neural networks (152). ATAC-sequencing is a very useful data modality for informing on this set of tasks (153).

More recently, researchers have adopted transformers and architectures similar to large language models, such as BERT, to segment regulatory regions and predict expression levels (53). These models are typically trained on a large number of DNA sequences or similar kinds of data using a supervised fashion by employing a masking strategy to generate a labeled dataset when labels are not available (53,148). This entails subdividing a DNA sequence into numerous smaller fragments, typically employing a k-mers based approach, where a k-mer is a subsequence of length k within the DNA. Selected k-mers are then masked within the sequence. The objective of the transformer-based model is to predict those masked parts of the sequence while imitating the interactions among different fragments (53,154). By predicting the masked fragment, the model learns the underlying structure of the DNA sequences. The resulting output is a set of numerical vectors, often referred to as embeddings, representing each k-mer, encapsulating the sequence information. These vector representations can be further used to fine-tune the model for various downstream tasks, such as segment identification, sequence alignment, and gene expression prediction where limited labeled data is available (53,148,149).

When labeled data is available, convolutional neural networks and transformers can be combined to predict outcomes. This approach employs convolutional neural networks on the small DNA fragments to obtain an initial representation. Following this, the processed fragments are inputted into transformer layers. The transformer architecture finds how each fragment interacts with other fragments and summarizes these interactions in a vector that represents the fragment. The vector representations of all fragments are finally used to predict different genomic tracks (155). Since the vector representations of the DNA fragments derived from the final transformer layer encapsulate information that is both comprehensive and adaptable across various genomic tracks, these vectors can be effectively utilized for predicting a range of genomic tasks beyond those initially targeted.

Since these approaches can be used for gene expression prediction, they are particularly useful for predicting the expression of an edited DNA sequence. Gene editing involves adding, removing, or substituting a segment of DNA to alter the expression of a specific gene. Identification of replacement DNA segments is a complex task that can be addressed by employing an ML algorithm that can estimate the gene expression. Alternatively, using RL, the discussed ML approaches can be trained to generate gene-edited DNA sequences that maximize the expression by following the same training architecture of Large Language Models, such as ChatGPT (156). Initially, one of these models can be fine-tuned to generate a gene-edited DNA sequence from an input DNA sequence (supervised fine-tuned model). Then, the supervised fine-tuned model can be optimized for predicting gene expression, which will work as the reward function of the RL agent (reward model).

Finally, the supervised fine-tuned model can be further optimized by employing RL optimization techniques, such as Proximal Policy Optimization (157), where outputs of the reward model are used as rewards for generated sequences.

## 4  Media Design

Cell culture requires media to keep cells alive, promote growth, and direct differentiation. Culture media contains the nutrients and signals that cells receive in the body and is typically composed of carbohydrates, amino acids, vitamins, minerals, buffers, proteins, peptides, fatty acids, lipids, and growth factors. In addition, media components may also contribute to the flavor of the final product (158,159). A detailed discussion of how cells utilize these components has been covered previously (158).

Most existing cell culture media formulations are expensive, not designed for the species of interest for CM production, and include animal-derived ingredients (158). Media design for CM requires optimizations to address these limitations, which are often complex and resource-intensive research efforts. ML is well suited to accelerate research in this space.

### 4.1  Culture Media Design Can be Treated as a Hyperparameter Optimization Problem

The objective of media design is to identify conditions that are optimized for various parameters, such as price, growth rate, stability, flavor, environmental impacts, and lot-to-lot reproducibility, sometimes optimizing for several of these objectives simultaneously (158,160). In addition, media may be optimized for specific cell responses, such as differentiation or spontaneous immortalization (161,162). In certain cases, media may need to be developed for uncharted parameter spaces; for instance, cells derived from lamb, duck, and deer that were not previously grown for biomedical purposes.

Most culture media contain dozens of ingredients - the classic media Ham's F-12 contains 47 ingredients (163). Certain ingredients, such as fetal bovine serum or B27 (164), are themselves complex mixtures of additional ingredients. In the absence of ML, single specific improvements of large effect can still be sometimes made to culture media, such as a group that swapped animal albumin for plant albumin (165). Big gains in the stability of stem cell media have been made with recombinant peptides (166). Other groups have optimized media, such as bovine myoblast media, using straightforward factorial approaches, and design of experiments for these approaches are quite mature (167,168). However, a medium with 49 ingredients, each with five relevant concentrations, has a design space of over 1034 potential recipes, exceeding the capacity of traditional factorial methods (i.e., Taguchi methods (169)) or high-throughput screening.

Some attempts have been made to apply ML to the culture media design problem as a means of exploring the experimental search space more effectively. A recent preprint (170) used gradient-boosted trees to identify culture media that improve the growth characteristics of suspension-phase HeLa cells. Another recent paper trained a neural network to model the results of fractional factorial culture media experiments (21). Response surface methodology, a classic design-of-experiments method, has been used in conjunction with a genetic algorithm for optimizing media properties such as cost, growth rate, and global warming potential of both zebrafish fibroblast (21) and mouse myocyte (20) culture media. However, these methods, on their own, aren't optimal for dealing with the complexity of full culture media recipe design.

Designing culture media is fundamentally a high-dimensional search problem, and Bayesian optimization can effectively navigate problems of this nature, which has the advantages of requiring only a small number of experiments and dealing with uncertainty in a robust manner. Bayesian optimization has been used to optimize myocyte culture media (18), spirulina culture media (171), and keratinocyte differentiation media (172). Google Vizier, which is Google's hyperparameter tuning service, is a key enabler of Bayesian optimization, as it provides biologists access to this algorithm via API, bypassing the need to code from scratch (173). Additionally, the BoTorch library for Python offers a powerful free toolkit to implement Bayesian optimization (174).

Even given an optimal algorithm for optimization, obtaining and analyzing relevant data from cultured cells remains a challenge in media design. One potential solution is through the use of high-content imaging, which can be enhanced through the implementation of ML methods, as discussed below in Section 5. Another approach involves the application of gene network analysis, which may be the most effective method for gaining detailed insights into the cellular response to a given culture medium.

Bayesian optimization is the leading hyperparameter exploration technique, but it stands among others. Genetic algorithms, which cluster parameters in a way that mimics biological chromosomes to enable recombination (175), have been used to optimize cyanobacteria media (176) and myocyte media (19). Simulated annealing, which is based on a metallurgical concept (177), has been used in conventional animal agriculture to optimize poultry feed (178) and to produce a population balance model for CHO cells (178). However, Bayesian optimization appears to be the most widely used among these techniques, probably because it can work on non-differentiable objective functions without the difficulty of partitioning parameter space into somewhat arbitrary and finicky "chromosomes" (179).

**4.2   Machine Learning Can Interpret How Cells Respond to Media Conditions**

The heart of media design is attempting to alter the properties of cells by altering their environment. Assessing the properties of cells can be challenging - although RNA-seq can provide insight on the inner state of cells by quantifying expressed genes (161). However, that information is often not readily interpretable because it is high-dimensional and the effect of any single gene is often esoteric or context-dependent. ML-guided network analysis is a solution - it can assess changes to the inner state of cells when performing media design or other optimizations, such as culture vessel geometry or atmosphere composition. Both gene-based network analysis and metabolite-based network analysis can be used for this application.

Gene network analysis has not yet been directly used in CM media research, but it has been used in parallel applications. Network analysis has been used to improve mammalian culture, CHO cells in particular, to identify feed supplements (180). Network analysis has also been used to find gene networks important for feed efficiency or milk production in cattle, with the networks formed from phenotype-genotype analysis (181,182).

To get more details about a gene network, gene network inference (GNI) methods can deduce the regulatory interactions between genes. The potential value of GNI to CM media design is that GNI can aid and simplify the interpretation of an RNA-seq data matrix; instead of looking at the differential expression of tens of thousands of genome features, the analyst can look at the differential activity of dozens of network modules. Among recent GNI methods are dynGENIE3 (random forests) (183), Scribe (184), and TENET (185), all of which are benchmarks for the most recent papers (186). GNI has been used to optimize output in traditional agriculture, both soybean cultivation (187) and maize cultivation (188), and GNI could be used for analogous output maximization problems in CM.

In biotechnology, metabolic network analysis has, historically, most commonly been done with linear programming. Flux balance analysis, a type of linear programming that is especially prominent, attempts to predict the inputs and outputs of metabolites in cells (189). Linear programming models can incorporate the relationships between different biological networks, assuming the relationships between networks are both known and linear (190,191). Flux balance analysis has been used to model CHO metabolic states (192), to predict the effect of sparging on cultures (193), to evaluate hypotheses regarding metabolic changes (194), and to develop genome-scale metabolic models (195), all of which are relevant aspects of CM. Linear programming has been combined with other ML methods in CHO cells (180) for optimizing cell growth.

## 4.3 Optimizing and Engineering Proteins Is Much Easier With Machine Learning

Another opportunity for optimization is in the ingredients themselves. Growth factors are currently estimated to contribute over 95% of the total production cost (196). Altering proteins for use in cell culture media is a promising means to optimize the stability, price, or other desirable properties of these ingredients.

Recombinant production, a commonly used animal-free strategy to produce proteins, is a major contributor to media costs (197). Most proteins present in culture medium are highly specialized eukaryotic proteins, requiring post-translational modifications such as phosphorylation or glycosylation that require mammalian cell culture systems, rather than the far cheaper and more scalable bacterial production systems. Recent work developed an E. coli expression system to reduce the price of several relevant growth factors (197). Similar efforts are necessary to reduce the cost of other media ingredients.

Probably the biggest challenges when importing transgenic proteins are differential folding between species (i.e. different chaperones, different organism temperatures) and differential post-translational modification. There have been many efforts to use ML to predict post-translational glycosylation, with earlier efforts using random forests (198) and more recent efforts using multi-layer perceptrons (199) or ensemble models (200). As for differential folding, transformer-based models have been used to predict thermal protein stability (201). Combining transformers and GNNs has been used to predict protein subcellular localization (202). More recently, AlphaFold has been used to predict protein stability in a general sense (203), and ColabFold was developed to improve upon it (204).

Protein ingredients can also be engineered to optimize their stability in culture, reducing the total amount of protein needed for CM production (205). For example, FGF2 thermal stability was increased using point mutations in the protein (206). Similarly, Long R3 IGF-1 is a modified recombinant form of IGF that prevents inactivation by IGF binding factors, leading to x200 potency and x3 stability compared to standard insulin (207). Similar strategies could be applied to other expensive media ingredients to reduce the necessary concentrations or additions. RFdiffusion (based on a generative diffusion model) (208) and ProteinMPNN (209) are tools designed to solve this problem. ML is also used in directed evolution experiments that iterate upon a given protein's design (142,210).

For some proteins used in high concentrations, such as albumin, transferrin, and insulin, replacement with lower-cost alternatives, such as plant (146,211) or microbial (8,212) hydrolysates may be beneficial. For example, a protein similar to bovine insulin was found in cowpea (Vigna unguiculata) and could be isolated for use as an insulin replacement (213,214). Screening plants or microbes for sequence homology to proteins of interest could accelerate this work. The traditional, non-ML tool for homology searches is protein BLAST (215), but BLAST is underpowered because proteins can have similar structures without having similar sequences. Services like Dali provide protein structure alignment but are constrained by the paucity of known protein structures and the computational complexity of three-dimensional comparisons (216). ML has changed how to approach this problem with FoldSeek (217), which uses autoencoder-based compression to simplify three-dimensional protein structural comparisons, using AlphaFold to infer protein structures when no empirical data is available (38).

## 5 Cell Culture Microscopy and Image Analysis

Microscopy is one of the foundational techniques of cell culture, providing information such as (i) the health of cells (e.g., whether they are mitotic, senescent, or apoptotic); (ii) the behavior of cells (e.g., whether they are invasive, contractile, or secretory); and (iii) the lineage of cells (e.g., whether they are stem cells, progenitor cells, or terminally differentiated cells). Many types of cell analysis rely on microscopy. For example, fusion index, the percentage of nuclei inside myotubes using immunostaining, is the predominant test used to quantify myogenic differentiation (64). Alternatively, inexpensive colorimetric staining has been shown as an alternative measure of myotube differentiation (218). Additionally, brightfield imaging, without the use of dyes, has been used to measure cell contractility, which is a measure of muscle cell maturity (219,220).

The relevance of microscopy during large-scale production will be especially dependent on its ability to serve as a low-cost, high-throughput tool. However, microscopy has been historically constrained by the complexity of its analysis. Microscopy analysis is routinely done manually by researchers with well-trained eyes, and doing it automatically requires systems that can incorporate many nuanced features of the image data. Unfortunately, appropriate systems tend to be nascent, poor, or non-existent in biological research. Furthermore, the use of dyes to improve image quality is undesirable due to the cost and time constraints of CM production. For CM production, cell segmentation and classification are fundamental and indispensable due to their multifaceted contributions - they are relevant to quality control, monitoring of cell culture health, and optimizing production. Utilizing ML approaches for automated cell segmentation and classification can lead to a reduction in the time, expenses, and errors involved in preparing the setup for manually analyzing the image data.

### 5.1 Automatic Image Analysis for Cell Segmentation using Machine Learning

Cell segmentation is the process of identifying and separating individual cells within an image. This is done in digital microscopy and histopathological imaging to study cell structure and function. The goal of cell segmentation is to pick out cells in an image, and segmentation is necessary to measure cell size, shape, and number. Segmentation also enables the tracking of individual cells over time, allowing researchers to assess changes in cell behavior and morphology, which can help in optimizing conditions for CM production. Furthermore, cell segmentation can also help in identifying contaminating debris or inappropriate cells from the culture, ensuring the quality and safety of the final product.

Recently ML has been used to improve cell segmentation for various applications (221–224). Manual identification of individual cells has poor reliability between and within human evaluators. Additionally, automating cell imaging tasks can free up researchers' time for higher value tasks and reduce errors from fatigue and subjectivity (225). However, there are several challenges that hinder the effectiveness of ML-based cell segmentation. First, cells are highly variable in size, shape, and morphology. Second, segmentation may fail on images that are low-contrast, have uneven illumination, or are out-of-focus. Third, crowded and overlapping cells can make it difficult to distinguish individual cells using cell-segmentation algorithms. Fourth, training data can overfit on imaging artifacts, such as non-uniform illumination and background noise (226). These challenges are not unique to CM but must be taken into account prior to applying ML for cell segmentation in any field.

In order to overcome the above challenges and to develop robust algorithms for detecting and segmenting cells, the computer vision community requires access to large, diverse, and well-curated datasets with comprehensive annotations. While some public datasets for nuclei and cell segmentation have been released in the past (223,225,227–230), additional datasets specific to CM are necessary to develop such robust algorithms for accurate cell segmentation in microscopic images.

A classic method for cell segmentation is the watershed algorithm combined with optimal thresholding, which has been used in applications such as segmenting lymphocytes into nuclear and cytoplasmic regions (231). There has been growing interest in deep neural net architectures inspired by fully convolutional networks for cell segmentation (223,230,232–235). These architectures employ encoder-decoder blocks to transfer features from multiple scales and levels for efficient cell segmentation on histopathology and microscopy images. The U-Net model, a variant of the fully convolutional network architecture, has shown particular promise for this task (236,237). Unlike fully convolutional networks, U-Net incorporates skip connections that facilitate precise semantic segmentation by amalgamating features from diverse resolutions, enhancing the model's capability to capture intricate details. Similarly, U-Net++ employs advanced encoder-decoder structures and loss functions to enhance performance for cell segmentation (238). Such deep learning architectures have outperformed pathologists' performance for cell segmentation in various applications (239,240).

## 5.2 Automatic Image Analysis for Cell Classification using Machine Learning

Cell classification is another important task in image analysis. While cell segmentation is the process of separating individual cells from the background and from each other in an image, cell classification is the process of assigning labels to the segmented cells based on their morphology, phenotype, or function (241,242). The primary goal of cell classification is to assign each segmented cell to a specific category or label, such as cell type, state, or condition.

Manual analysis of thousands of microscopy images for cell classification is a tedious and error-prone task. Thus, there is a need for ML algorithms for automatic cell classification. Recently, ML has been used for the classification of cells within microscopy images, including the identification of anemia or blood disorders based on the shape, size, and optical properties of red blood cells (243), and for the analysis of cellular microenvironments to offer novel insights into biological mechanisms (244). However, the effectiveness of ML-based cell classification for microscopy faces several challenges. One challenge arises from the crowded or overlapping cells, making it difficult to distinguish individual cells. Differences in cell maturity and variations in cell shape resulting from diverse cultivation and treatment methods can introduce complexity into the analysis, especially when distinguishing between different types of cultured cells. Complex textures, patterns, and shapes in multi-modal microscopy further complicate cell classification, requiring the integration of multi-stream models that consider low-level cues such as edges and gradients (245). The intricate cell structures found in tissue images introduce additional complexities such as heterogeneous cell populations and staining variations. These challenges necessitate the development of precise and efficient algorithms for cell classification.

In the domain of microscopy image analysis, many innovative architectures and methods have emerged for classifying cells. Among these, CNNs stand as a robust choice, with models like U-Net (237) and Mask R-CNN (246) excelling in cell classification tasks. RNNs, particularly long short-term memory and gated recurrent units, demonstrate their prowess when handling sequential data, making them valuable for tracking cell dynamics (247). For tasks involving complex relationships between cells, GNNs, such as graph convolutional networks, prove invaluable (248). Traditional approaches like random forests and decision trees, which are based on hand-picked image features, are still relevant for classification because they are computationally fast, easy to train, and resist overfitting (249,250). Transfer learning techniques harness pre-trained deep learning models like ResNet (251), while attention mechanisms and ensemble methods contribute to improved accuracy (252). With the ever-evolving landscape of microscopy, these versatile architectures continue to play pivotal roles in cell classification.

Following cell classification, cell phenotype analysis is pivotal in exploring cellular characteristics, encompassing physical and biochemical attributes such as size, shape, function, viability, proliferation, signaling, and morphological structure. This approach is particularly valuable in culture media optimization (253). By scrutinizing the physical attributes and functional behavior of cultured cells, researchers can ensure the consistency and quality of cell populations, evaluate metabolic activity, and assess cellular functionality within the culture. Moreover, the examination of cell morphology offers insights into cellular health and enables fine-tuning of culture media formulations (253,254). ML has been used to combine different forms of microscopy using transfer learning - such as between fluorescence and dye-free microscopy (255) or between traditional microscopy and mass spectrometry imaging (256). Thus, in the burgeoning field of CM, phenotype analysis contributes to the refinement of CM production, guiding the selection of cell strains and culture conditions to achieve better quality and desired growth attributes. Overall, cell phenotype analysis plays a pivotal role in enhancing cell culture processes and product quality, making it a valuable tool for CM production.

## 6      Bioprocess and Food Processing Optimization

Moving CM from lab bench scale to commercial scale requires efficient bioprocess design. This centers around the use of large bioreactors to produce a controlled environment for cell growth and differentiation that maximizes biomass and minimizes by-product yields. A variety of bioreactor types have been proposed for use in CM, which are reviewed elsewhere (257). One study estimated that producing 1kg of protein from muscle cells would require stirred tank bioreactors on the order of 5000 liters (258). This dwarfs research-scale mammalian cell culture and will require extensive optimization. In addition, after harvesting cells from the bioreactor, CM products will likely require food processing steps to create a final product. ML is well suited to increase the scale and efficiency of CM bioprocessing and food processing in a variety of ways.

### 6.1     Bioreactor Homeostasis Can be Maintained with Machine Learning

As the culture scale is increased, automation of closed systems will become important to increase efficiency and reduce contamination or other failure events (259). Real-time quality assurance and process monitoring will allow for adjusting culture conditions to optimize yield, monitoring for potential contamination, and reducing human errors. Automation in processes such as media recycling will further optimize the process and bring costs down.

Historically, mammalian bioreactor operation has been governed by systems such as proportional-integral-derivative controllers (260) or model predictive control (261). These systems, although mainstays of control theory, are designed to work in deterministic environments that can be well-described by linear differential equations. Although these systems can struggle with the vagaries and unpredictability of biological systems, they have been successfully employed in bioreactors to maximize antibody production (262), constrain overflow metabolism (263), and maintain glucose homeostasis (264). This type of modeling has ceded ground to ML-based modeling in recent years, probably because, compared to ML, these models are relatively fragile because of their dependence on their top-down mathematical models accurately describing reality.

The most straightforward application of ML to bioreactors is to use ML-based models to control the bioreactor's inputs. A wide gamut of ML models have been used to monitor and control bioreactors and industrial bioprocesses, especially supervised learning models like neural networks (265), random forests (266), and gradient boosting (267). The step beyond using ML to optimize models is to use ML to optimize the policies themselves that determine the models, which is done by reinforcement learning, with examples in bioreactors ranging across deep Q-networks (268), policy gradients (269), and probabilistic Bayesian optimization (270). These policy-based learning methods do not necessarily require a prior understanding of the biochemistry of the bioreactor.

**6.2     The Unique Challenges of Structured Products Can Be Addressed by Machine Learning**

As opposed to ground meat, structured tissues (i.e. a steak or fish filet) will require the formation of organized 3-dimensional tissues and bioreactor systems capable of supporting them. A structured product entails growing an organized and three-dimensional tissue, as opposed to growing cells in suspension - the difference between growing cells as a soup and growing cells as a steak. This type of engineered tissue has not been shown on any scale beyond a tissue for a single patient, making this a major whitespace in the field (259).

Imposing organization upon cells is a classic, challenging problem from the field of tissue engineering, and solutions often involve ML. Random forests (271) and neural networks (272) have been used to predict optimal parameters for the extrusion printing of hydrogel scaffolds. Gradient boosting has been used to predict the self-assembly of dipeptide-based hydrogel scaffolds (273). In theory, ML could be used to design the structures of tissue scaffolds, but this avenue appears to be unexplored at this time.

Tissue self-organization is a powerful principle for structured meat production. Tissue scaffolds or extrusion printing are currently used to create top-down structured tissues because of a lack of understanding of bottom-up self-organization of tissue in vitro. However, breakthroughs in bottom-up tissue self-organization would accelerate the scale-up of CM, eliminating the costs of scaffolds. In the context of CM, relevant types of self-organization are the alignment of fibers (especially myofibrils and collagen cables), the production of functional vasculature networks, maintaining ratios of meat-relevant cell types (myocytes, adipocytes, fibroblasts), and the structural determinants of texture and mouthfeel (274). Attempts have been made to model tissue self-organization using differential adhesion (275), cellular Potts models (276), and agent-based modeling (277). Accurate models of tissue self-organization may provide design principles for making tissues with defined structure in CM.

A family of methods that may help with structured tissue construction in CM is spatial transcriptomics, which correlates gene expression in situ to physical coordinates within a tissue section (278). Spatial transcriptomics has been used for understanding specific aspects of tissue structure and cell organization that rely on spatial context, such as extracellular forces and gradients of signaling molecules (136). GNNs have been successfully used with spatial transcriptomics data to model cellular communication (279–281) and the deep learning model Tangram has also been used to resolve cell types and decrease imputation error from spatial data (282). Although spatial transcriptomics has not yet been used in the context of CM, it could prove useful in dissecting the complex tissue architectures involved in structured products.

Structured products will also require bioreactor systems capable of perfusion and harvesting of large intact tissues. The nature of a structured product bioreactor bears similarity to the fluidized and/or packed bed bioreactors that are used industrially in the wastewater and mineral extraction industries. Numerical fluid dynamical models are classically used to predict the behavior of these systems, but ML has, in recent years, been used for the more unpredictable aspects of these systems (283,284). In particular, gradient boosting has been used both to predict bed expansion of fluidized bed bioreactors (285) and mass transfer in packed bed bioreactors (286). Similar methods may be useful for the particular challenge of CM structured product bioreactors.

### 6.3 Real-Time Sensory Prediction and Control Could be Applied with Reinforcement Learning

Unlike tissue engineering for medical treatments, the sensory properties of CM, such as flavor and texture, are critical to its commercial success. These may be generated during the cell culture from flavor or texture components of cells, media, or scaffolds, or after harvest using food processing or additives. ML is playing an important role in enhancing the analysis of the flavors and textures of other food products through the analysis of diverse data types and these techniques are likely to play a role in CM development as well.

During product development, flavor can be measured in a variety of ways. Earlier ML models used data from gas chromatography-mass spectrometry, which is an analytical chemistry method used to separate and fingerprint substances from complex mixtures (287,288). Later, researchers focused on electronic noses that mimic the olfactory capability of humans through different sensors. Since an electronic nose can be used to collect real-time sensor data during production, ML can also be applied in real-time for quality and flavor control (289–291). However, it should be taken into consideration that many compounds important for the aroma of meat are generated during cooking (292). The latest research efforts concentrate on utilizing the molecular structure and physicochemical attributes of flavor compounds to predict flavors, including taste or smell (293–297). These characteristics are quantified into molecular descriptors, numerical representations that encapsulate the properties of the molecules involved. These descriptors then serve as inputs for ML models, which are trained to predict flavor profiles and odor characteristics with greater objectivity.

Since most of these data are tabular, a wide range of traditional ML approaches, such as support vector machines, random forests, k-nearest neighbour, and AdaBoost Tree, have been employed (295,297,298). Additionally, deep learning approaches, such as CNNs (287) and multilayer perceptrons (288,289,291), and unsupervised learning approaches, such as cluster analysis by using principal component analysis (289,291), have also been applied to identify or predict flavors. These methodologies have collectively demonstrated that ML can significantly contribute to the enhancement of the sensory properties of a range of food products.

AI approaches, particularly RL, offer significant potential for enhancing efficiency in food processing operations (269,299,300). Food-processing facilities are typically equipped with a variety of sensors, including those for temperature, pressure, moisture, and pH levels. These sensors play a crucial role in ensuring precise ingredient measurements, which are fundamental for achieving the desired taste and aroma profiles of food products. However, the challenge arises when new products are developed, as determining the optimal mixture of ingredients often involves extensive trial and error. Deploying an RL agent in this context can effectively manage and adjust the various sensor readings, thereby ensuring that the food consistently meets the specific standards and requirements set by the food processor. This approach not only streamlines the development process but also enhances the precision and quality of the final product. RL can also be valuable for sensory prediction, in particular the texture prediction of finished CM products (301).

# 7 Discussion

In the last decade, the field of CM has made strides toward lower-cost and more efficient production processes but must progress significantly further to effectively rival traditional meat. ML offers great promise in improving every stage of CM production, from cell line development to the final product's sensory characteristics. With the current surge in both public and private research and development for both the ML and CM fields, and the already successful integration of ML into numerous life sciences fields, the integration of ML into CM research is timely. Despite the numerous opportunities, there are only a handful of peer-reviewed, publicly accessible studies that describe the use of ML in CM production, and an equally small group of researchers versed in both ML and CM. This review aims to bridge the gap between the ML and CM fields and create a starting point for scientists to better understand how to apply ML to CM research.

Since ML has been successfully employed in many other bioinformatics sectors, many of the existing methods can be adopted to CM. However, the key limitation is the availability of sufficient data. Creating ML models requires large training and validation datasets, and ML models are only as robust and reliable as the amount and quality of data used to develop them. The scarcity of public data in CM complicates the development of models or even the assessment of potential model types. There are some publicly available dataset repositories, such as the Gene Expression Omnibus (GEO) for RNA-seq data[1], GenBank for sequenced genome data[2], and Uniprot for protein sequences[3]. However, compared to species used in medical studies, few CM-relevant datasets are reported and many lack adequate descriptions, data annotations, or samples and replicates to be considered for statistical analysis. As an example, a query for "stem cell" in the GEO DataSets generates 71327 datasets for Homo sapiens (human) and only 61 for Bos taurus (cattle) (as of Feb 1, 2024). A survey of datasets with potential relevance to CM research is included in Supplementary Table 1 (hosted on Zenodo, DOI: 10.5281/zenodo.11092402). Efforts to generate properly annotated data and incentives for the sharing of data from ongoing experiments would greatly accelerate the application of ML to CM research. The Cultivated Meat Modeling Consortium[4] offers a model for community-generated and shared data, while protecting intellectual property, to accelerate computational models.

Transfer learning might be able to partially make up for the scarcity of CM-relevant data, by using models from data-rich biomedical species, such as humans and mice, to inform models for data-poor CM-relevant species. Cross-species graph-based transfer learning has previously been applied in a non-CM context on RNA-seq data for cell-type identification (302,303). Challenges to this approach include biological heterogeneity from differing sets of genes or differing functions for genes across species, which can potentially be mitigated with gene orthology information (304). As a starting point, researchers may look into fine-tuning large scale models that have been successful in achieving improved performance on biological tasks relevant to CM with limited task-specific data, including gene network analysis and cell type annotation, such as Geneformer and scGPT (305,306).

---

[1] https://www.ncbi.nlm.nih.gov/geo/
[2] https://www.ncbi.nlm.nih.gov/genbank/
[3] https://www.uniprot.org/
[4] https://thecmmc.org/

Another challenge relates to the scale of existing studies. Most ML research related to CM has been limited to laboratory environments, which might not mirror the conditions of mass production. Solutions could include either verifying these lab models at a commercial level or using process simulations to adapt them for larger operations.

From the perspective of biologists, an important way to speed up ML work is to produce the biological models that are used to generate data for training ML models. For example, the production and dissemination of more high-quality cell lines from agriculturally relevant animals is needed to aid the generation of omics and microscopy data. Furthermore, biological scientists can make efforts to contribute to the body of existing data by publishing any quality data that results from their experiments, whether or not they are directly used in their own studies, including negative results. Publication of transcriptomic and epigenomic data has become more commonplace, however complete microscopy image sets, in particular, are rarely published. There is also a need for more proteomic and metabolomic data, especially in understudied food-relevant species, to address numerous use cases including metabolic modeling, flavor profiling, and bioreactor scaling concerns (307). Moreover, datasets should be properly annotated - this first includes metadata describing the samples the data came from, how the data was generated, and any processing that was done to the data. Ideally, data would include all raw data, which ML scientists could use to regenerate the original dataset. Similarly, properly labeled data (labeling of individual data points) is important for supervised learning models, which remain the most popular and widely used. Finally, the quality and consistency of datasets are also critical for their use in training ML models. Variations in sample handling and data collection can lead to large and varied systematic errors that make it challenging to usefully combine multiple datasets or for a model trained on one dataset to apply to another.

Efforts to make ML more accessible to researchers, such as infrastructure, frameworks, benchmarks, and libraries, could help facilitate the application of ML to CM. Currently, a variety of ML models are freely available from online resources like Paperswithcode[5] and Hugging Face[6]. Additionally, libraries such as Python's scikit-learn (308) or frameworks such as PyTorch Lightning[7] offer a good starting point for coding ML. Similarly, accessible web interfaces could make ML tools more accessible to biologists, following the examples of AlphaFold and Foldseek, which both have interfaces integrated into the widely used online protein database UniProt.

Overall, there is an enormous opportunity for CM researchers to incorporate ML techniques and for ML professionals to explore the CM field. The ML field has been moving at unprecedented speed over the past few years, and CM researchers could make gains simply by porting over what is, in effect, yesterday's news in ML. This review aims to be an introductory resource for researchers eager to explore this cross-disciplinary method, which could help to establish CM as a viable and sustainable protein source in our diets.

---

[5] https://paperswithcode.com/
[6] https://huggingface.co/
[7] https://lightning.ai/

**Abbreviations**

AI: artificial intelligence

CHO: Chinese hamster ovary

CM: cultured meat

CNN: convolutional neural network

FAO: Food and Agriculture Organization

GAN: generative adversarial network

GNI: gene network inference

GNN: graph neural network

ML: machine learning

NLP: natural language processing

OECD: Organization for Economic Co-operation

RL: reinforcement learning

RNN: recurrent neural network

VAE: Variational Autoencoder

**Conflict of Interest**

The authors declare that the research was conducted in the absence of any commercial or financial relationships that could be construed as a potential conflict of interest.

**Funding**

BD, MET, JS, and RV were funded by a grant from the Schmidt Futures.

**Acknowledgments**

The authors would like to thank Zachary Consenza, Roy Nadler, Amin Nikkhah, Dave Staszak, Charlie Taylor, Sofia Giampaoli, Animesh Acharjee, Ana Velez Rueda, Jessica Carballido, and Leandro Sommese for reviewing early drafts of this manuscript and providing valuable insights. We thank Isha Datar and Cam Linke for their assistance in finding funding for the project. We would also like to thank Evan Rapoport for his facilitative leadership on this project.

**Figures**

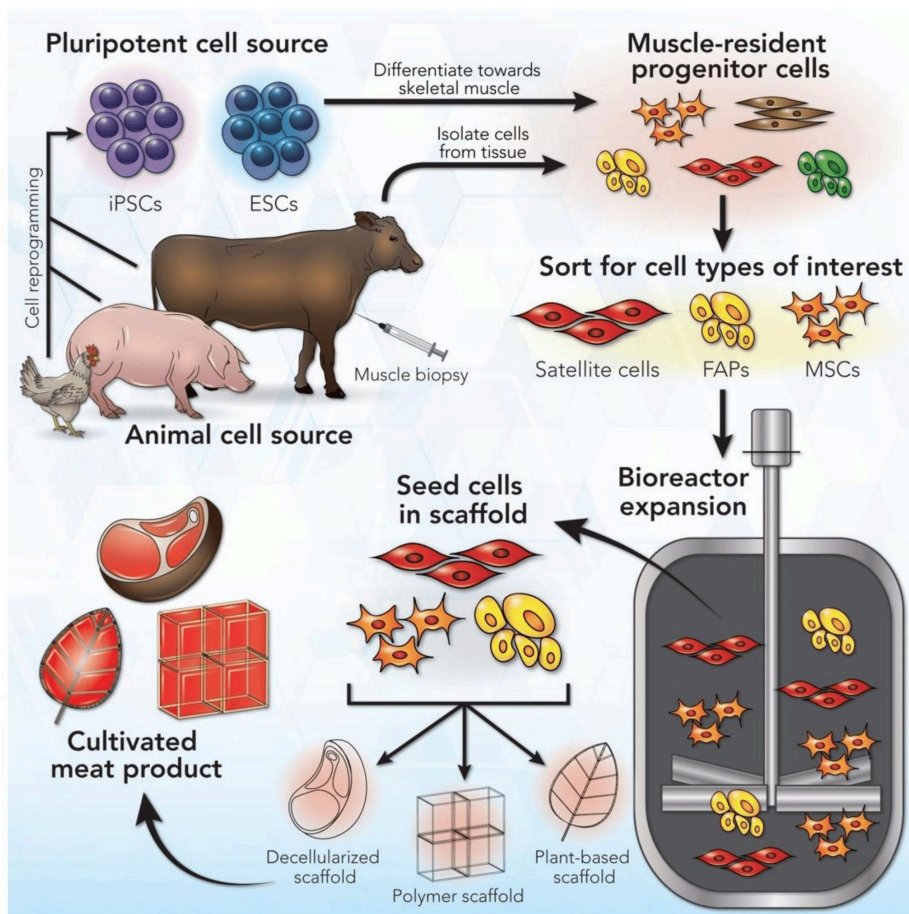

**Figure 1.** Cultured meat manufacturing process. Reprinted from Reiss, J. et.al., 2021. Creative Commons Attribution (CC BY) license. (29)
21

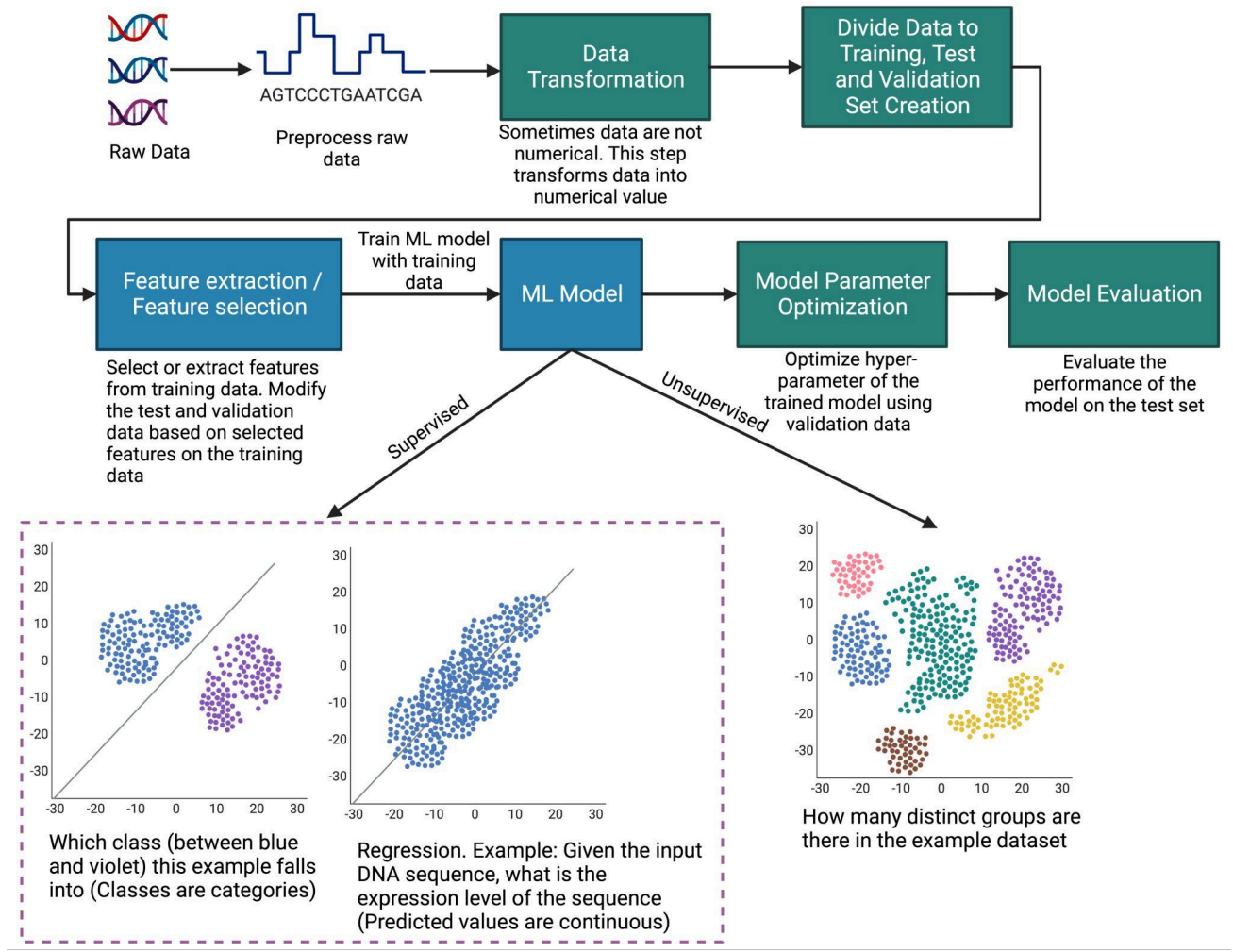

**Figure 2.** Illustration of the typical steps of how machine learning can be applied to biological data.

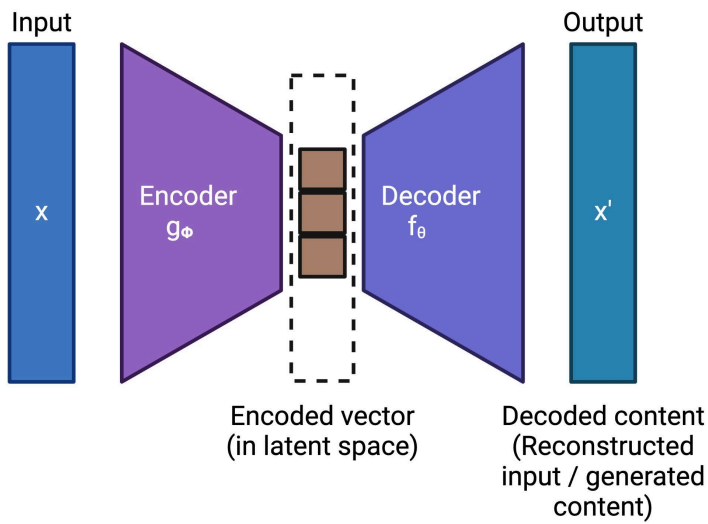

**Figure 3.** General architecture of a Variational Autoencoder (VAE).

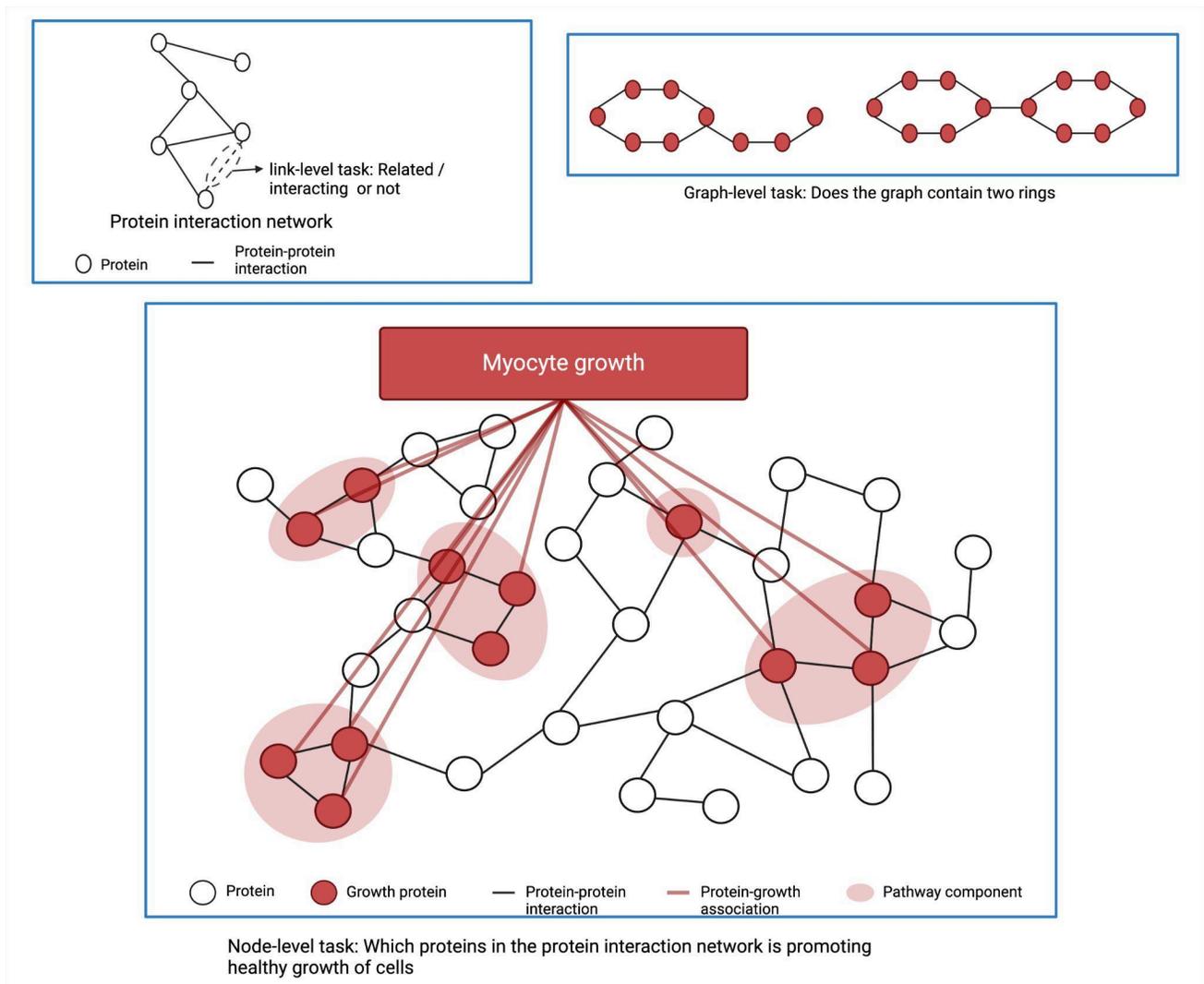

**Figure 4.** Typical machine learning tasks for network analysis.